\documentclass[sigconf, screen]{acmart}
\AtBeginDocument{%
  \providecommand\BibTeX{{%
    \normalfont B\kern-0.5em{\scshape i\kern-0.25em b}\kern-0.8em\TeX}}}

\copyrightyear{2024}
\acmYear{2024}
\setcopyright{rightsretained}
\acmConference[SIGIR '24]{Proceedings of the 47th International ACM SIGIR Conference on Research and Development in Information Retrieval}{July 14--18, 2024}{Washington, DC, USA}
\acmBooktitle{Proceedings of the 47th International ACM SIGIR Conference on Research and Development in Information Retrieval (SIGIR '24), July 14--18, 2024, Washington, DC, USA}
\acmDOI{10.1145/3626772.3657662}
\acmISBN{979-8-4007-0431-4/24/07}

\makeatletter
\gdef\@copyrightpermission{
 \begin{minipage}{0.3\columnwidth}
  \href{https://creativecommons.org/licenses/by/4.0/}{\includegraphics[width=0.90\textwidth]{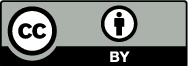}}
 \end{minipage}\hfill
 \begin{minipage}{0.7\columnwidth}
  \href{https://creativecommons.org/licenses/by/4.0/}{This work is licensed under a Creative Commons Attribution International 4.0 License.}
 \end{minipage}
 \vspace{5pt}
}
\makeatother

\usepackage{subcaption}
\usepackage{enumitem}
\usepackage{rotating}
\usepackage{microtype}
\usepackage{bm}
\usepackage{soul}
\usepackage{wrapfig}

\usepackage[varqu]{zi4}
\usepackage[utf8x]{inputenc}
\usepackage{units}
\usepackage{booktabs}
\usepackage{colortbl}
\usepackage{listings}

\usepackage{amsmath}
\usepackage{mathtools}
\usepackage{mathtools}

\definecolor{redOV}{RGB}{255, 235, 238}
\definecolor{redI}{RGB}{255, 205, 210}
\definecolor{redII}{RGB}{239, 154, 154}
\definecolor{redIII}{RGB}{229, 115, 115}
\definecolor{redIV}{RGB}{239, 83, 80}
\definecolor{redV}{RGB}{244, 67, 54}
\definecolor{redVI}{RGB}{229, 57, 53}
\definecolor{redVII}{RGB}{211, 47, 47}
\definecolor{redVIII}{RGB}{198, 40, 40}
\definecolor{redIX}{RGB}{183, 28, 28}
\definecolor{redAI}{RGB}{255, 138, 128}
\definecolor{redAII}{RGB}{255, 82, 82}
\definecolor{redAIV}{RGB}{255, 23, 68}
\definecolor{redAVII}{RGB}{213, 0, 0}

\definecolor{pinkOV}{RGB}{252, 228, 236}
\definecolor{pinkI}{RGB}{248, 187, 208}
\definecolor{pinkII}{RGB}{244, 143, 177}
\definecolor{pinkIII}{RGB}{240, 98, 146}
\definecolor{pinkIV}{RGB}{236, 64, 122}
\definecolor{pinkV}{RGB}{233, 30, 99}
\definecolor{pinkVI}{RGB}{216, 27, 96}
\definecolor{pinkVII}{RGB}{194, 24, 91}
\definecolor{pinkVIII}{RGB}{173, 20, 87}
\definecolor{pinkIX}{RGB}{136, 14, 79}
\definecolor{pinkAI}{RGB}{255, 128, 171}
\definecolor{pinkAII}{RGB}{255, 64, 129}
\definecolor{pinkAIV}{RGB}{245, 0, 87}
\definecolor{pinkAVII}{RGB}{197, 17, 98}

\definecolor{purpleOV}{RGB}{243, 229, 245}
\definecolor{purpleI}{RGB}{225, 190, 231}
\definecolor{purpleII}{RGB}{206, 147, 216}
\definecolor{purpleIII}{RGB}{186, 104, 200}
\definecolor{purpleIV}{RGB}{171, 71, 188}
\definecolor{purpleV}{RGB}{156, 39, 176}
\definecolor{purpleVI}{RGB}{142, 36, 170}
\definecolor{purpleVII}{RGB}{123, 31, 162}
\definecolor{purpleVIII}{RGB}{106, 27, 154}
\definecolor{purpleIX}{RGB}{74, 20, 140}
\definecolor{purpleAI}{RGB}{234, 128, 252}
\definecolor{purpleAII}{RGB}{224, 64, 251}
\definecolor{purpleAIV}{RGB}{213, 0, 249}
\definecolor{purpleAVII}{RGB}{170, 0, 255}

\definecolor{deeppurpleOV}{RGB}{237, 231, 246}
\definecolor{deeppurpleI}{RGB}{209, 196, 233}
\definecolor{deeppurpleII}{RGB}{179, 157, 219}
\definecolor{deeppurpleIII}{RGB}{149, 117, 205}
\definecolor{deeppurpleIV}{RGB}{126, 87, 194}
\definecolor{deeppurpleV}{RGB}{103, 58, 183}
\definecolor{deeppurpleVI}{RGB}{94, 53, 177}
\definecolor{deeppurpleVII}{RGB}{81, 45, 168}
\definecolor{deeppurpleVIII}{RGB}{69, 39, 160}
\definecolor{deeppurpleIX}{RGB}{49, 27, 146}
\definecolor{deeppurpleAI}{RGB}{179, 136, 255}
\definecolor{deeppurpleAII}{RGB}{124, 77, 255}
\definecolor{deeppurpleAIV}{RGB}{101, 31, 255}
\definecolor{deeppurpleAVII}{RGB}{98, 0, 234}

\definecolor{indigoOV}{RGB}{232, 234, 246}
\definecolor{indigoI}{RGB}{197, 202, 233}
\definecolor{indigoII}{RGB}{159, 168, 218}
\definecolor{indigoIII}{RGB}{121, 134, 203}
\definecolor{indigoIV}{RGB}{92, 107, 192}
\definecolor{indigoV}{RGB}{63, 81, 181}
\definecolor{indigoVI}{RGB}{57, 73, 171}
\definecolor{indigoVII}{RGB}{48, 63, 159}
\definecolor{indigoVIII}{RGB}{40, 53, 147}
\definecolor{indigoIX}{RGB}{26, 35, 126}
\definecolor{indigoAI}{RGB}{140, 158, 255}
\definecolor{indigoAII}{RGB}{83, 109, 254}
\definecolor{indigoAIV}{RGB}{61, 90, 254}
\definecolor{indigoAVII}{RGB}{48, 79, 254}

\definecolor{blueOV}{RGB}{227, 242, 253}
\definecolor{blueI}{RGB}{187, 222, 251}
\definecolor{blueII}{RGB}{144, 202, 249}
\definecolor{blueIII}{RGB}{100, 181, 246}
\definecolor{blueIV}{RGB}{66, 165, 245}
\definecolor{blueV}{RGB}{33, 150, 243}
\definecolor{blueVI}{RGB}{30, 136, 229}
\definecolor{blueVII}{RGB}{25, 118, 210}
\definecolor{blueVIII}{RGB}{21, 101, 192}
\definecolor{blueIX}{RGB}{13, 71, 161}
\definecolor{blueAI}{RGB}{130, 177, 255}
\definecolor{blueAII}{RGB}{68, 138, 255}
\definecolor{blueAIV}{RGB}{41, 121, 255}
\definecolor{blueAVII}{RGB}{41, 98, 255}

\definecolor{lightblueOV}{RGB}{225, 245, 254}
\definecolor{lightblueI}{RGB}{179, 229, 252}
\definecolor{lightblueII}{RGB}{129, 212, 250}
\definecolor{lightblueIII}{RGB}{79, 195, 247}
\definecolor{lightblueIV}{RGB}{41, 182, 246}
\definecolor{lightblueV}{RGB}{3, 169, 244}
\definecolor{lightblueVI}{RGB}{3, 155, 229}
\definecolor{lightblueVII}{RGB}{2, 136, 209}
\definecolor{lightblueVIII}{RGB}{2, 119, 189}
\definecolor{lightblueIX}{RGB}{1, 87, 155}
\definecolor{lightblueAI}{RGB}{128, 216, 255}
\definecolor{lightblueAII}{RGB}{64, 196, 255}
\definecolor{lightblueAIV}{RGB}{0, 176, 255}
\definecolor{lightblueAVII}{RGB}{0, 145, 234}

\definecolor{cyanOV}{RGB}{224, 247, 250}
\definecolor{cyanI}{RGB}{178, 235, 242}
\definecolor{cyanII}{RGB}{128, 222, 234}
\definecolor{cyanIII}{RGB}{77, 208, 225}
\definecolor{cyanIV}{RGB}{38, 198, 218}
\definecolor{cyanV}{RGB}{0, 188, 212}
\definecolor{cyanVI}{RGB}{0, 172, 193}
\definecolor{cyanVII}{RGB}{0, 151, 167}
\definecolor{cyanVIII}{RGB}{0, 131, 143}
\definecolor{cyanIX}{RGB}{0, 96, 100}
\definecolor{cyanAI}{RGB}{132, 255, 255}
\definecolor{cyanAII}{RGB}{24, 255, 255}
\definecolor{cyanAIV}{RGB}{0, 229, 255}
\definecolor{cyanAVII}{RGB}{0, 184, 212}

\definecolor{tealOV}{RGB}{224, 242, 241}
\definecolor{tealI}{RGB}{178, 223, 219}
\definecolor{tealII}{RGB}{128, 203, 196}
\definecolor{tealIII}{RGB}{77, 182, 172}
\definecolor{tealIV}{RGB}{38, 166, 154}
\definecolor{tealV}{RGB}{0, 150, 136}
\definecolor{tealVI}{RGB}{0, 137, 123}
\definecolor{tealVII}{RGB}{0, 121, 107}
\definecolor{tealVIII}{RGB}{0, 105, 92}
\definecolor{tealIX}{RGB}{0, 77, 64}
\definecolor{tealAI}{RGB}{167, 255, 235}
\definecolor{tealAII}{RGB}{100, 255, 218}
\definecolor{tealAIV}{RGB}{29, 233, 182}
\definecolor{tealAVII}{RGB}{0, 191, 165}

\definecolor{greenOV}{RGB}{232, 245, 233}
\definecolor{greenI}{RGB}{200, 230, 201}
\definecolor{greenII}{RGB}{165, 214, 167}
\definecolor{greenIII}{RGB}{129, 199, 132}
\definecolor{greenIV}{RGB}{102, 187, 106}
\definecolor{greenV}{RGB}{76, 175, 80}
\definecolor{greenVI}{RGB}{67, 160, 71}
\definecolor{greenVII}{RGB}{56, 142, 60}
\definecolor{greenVIII}{RGB}{46, 125, 50}
\definecolor{greenIX}{RGB}{27, 94, 32}
\definecolor{greenAI}{RGB}{185, 246, 202}
\definecolor{greenAII}{RGB}{105, 240, 174}
\definecolor{greenAIV}{RGB}{0, 230, 118}
\definecolor{greenAVII}{RGB}{0, 200, 83}

\definecolor{lightgreenOV}{RGB}{241, 248, 233}
\definecolor{lightgreenI}{RGB}{220, 237, 200}
\definecolor{lightgreenII}{RGB}{197, 225, 165}
\definecolor{lightgreenIII}{RGB}{174, 213, 129}
\definecolor{lightgreenIV}{RGB}{156, 204, 101}
\definecolor{lightgreenV}{RGB}{139, 195, 74}
\definecolor{lightgreenVI}{RGB}{124, 179, 66}
\definecolor{lightgreenVII}{RGB}{104, 159, 56}
\definecolor{lightgreenVIII}{RGB}{85, 139, 47}
\definecolor{lightgreenIX}{RGB}{51, 105, 30}
\definecolor{lightgreenAI}{RGB}{204, 255, 144}
\definecolor{lightgreenAII}{RGB}{178, 255, 89}
\definecolor{lightgreenAIV}{RGB}{118, 255, 3}
\definecolor{lightgreenAVII}{RGB}{100, 221, 23}

\definecolor{limeOV}{RGB}{249, 251, 231}
\definecolor{limeI}{RGB}{240, 244, 195}
\definecolor{limeII}{RGB}{230, 238, 156}
\definecolor{limeIII}{RGB}{220, 231, 117}
\definecolor{limeIV}{RGB}{212, 225, 87}
\definecolor{limeV}{RGB}{205, 220, 57}
\definecolor{limeVI}{RGB}{192, 202, 51}
\definecolor{limeVII}{RGB}{175, 180, 43}
\definecolor{limeVIII}{RGB}{158, 157, 36}
\definecolor{limeIX}{RGB}{130, 119, 23}
\definecolor{limeAI}{RGB}{244, 255, 129}
\definecolor{limeAII}{RGB}{238, 255, 65}
\definecolor{limeAIV}{RGB}{198, 255, 0}
\definecolor{limeAVII}{RGB}{174, 234, 0}

\definecolor{yellowOV}{RGB}{255, 253, 231}
\definecolor{yellowI}{RGB}{255, 249, 196}
\definecolor{yellowII}{RGB}{255, 245, 157}
\definecolor{yellowIII}{RGB}{255, 241, 118}
\definecolor{yellowIV}{RGB}{255, 238, 88}
\definecolor{yellowV}{RGB}{255, 235, 59}
\definecolor{yellowVI}{RGB}{253, 216, 53}
\definecolor{yellowVII}{RGB}{251, 192, 45}
\definecolor{yellowVIII}{RGB}{249, 168, 37}
\definecolor{yellowIX}{RGB}{245, 127, 23}
\definecolor{yellowAI}{RGB}{255, 255, 141}
\definecolor{yellowAII}{RGB}{255, 255, 0}
\definecolor{yellowAIV}{RGB}{255, 234, 0}
\definecolor{yellowAVII}{RGB}{255, 214, 0}

\definecolor{amberOV}{RGB}{255, 248, 225}
\definecolor{amberI}{RGB}{255, 236, 179}
\definecolor{amberII}{RGB}{255, 224, 130}
\definecolor{amberIII}{RGB}{255, 213, 79}
\definecolor{amberIV}{RGB}{255, 202, 40}
\definecolor{amberV}{RGB}{255, 193, 7}
\definecolor{amberVI}{RGB}{255, 179, 0}
\definecolor{amberVII}{RGB}{255, 160, 0}
\definecolor{amberVIII}{RGB}{255, 143, 0}
\definecolor{amberIX}{RGB}{255, 111, 0}
\definecolor{amberAI}{RGB}{255, 229, 127}
\definecolor{amberAII}{RGB}{255, 215, 64}
\definecolor{amberAIV}{RGB}{255, 196, 0}
\definecolor{amberAVII}{RGB}{255, 171, 0}

\definecolor{orangeOV}{RGB}{255, 243, 224}
\definecolor{orangeI}{RGB}{255, 224, 178}
\definecolor{orangeII}{RGB}{255, 204, 128}
\definecolor{orangeIII}{RGB}{255, 183, 77}
\definecolor{orangeIV}{RGB}{255, 167, 38}
\definecolor{orangeV}{RGB}{255, 152, 0}
\definecolor{orangeVI}{RGB}{251, 140, 0}
\definecolor{orangeVII}{RGB}{245, 124, 0}
\definecolor{orangeVIII}{RGB}{239, 108, 0}
\definecolor{orangeIX}{RGB}{230, 81, 0}
\definecolor{orangeAI}{RGB}{255, 209, 128}
\definecolor{orangeAII}{RGB}{255, 171, 64}
\definecolor{orangeAIV}{RGB}{255, 145, 0}
\definecolor{orangeAVII}{RGB}{255, 109, 0}

\definecolor{deeporangeOV}{RGB}{251, 233, 231}
\definecolor{deeporangeI}{RGB}{255, 204, 188}
\definecolor{deeporangeII}{RGB}{255, 171, 145}
\definecolor{deeporangeIII}{RGB}{255, 138, 101}
\definecolor{deeporangeIV}{RGB}{255, 112, 67}
\definecolor{deeporangeV}{RGB}{255, 87, 34}
\definecolor{deeporangeVI}{RGB}{244, 81, 30}
\definecolor{deeporangeVII}{RGB}{230, 74, 25}
\definecolor{deeporangeVIII}{RGB}{216, 67, 21}
\definecolor{deeporangeIX}{RGB}{191, 54, 12}
\definecolor{deeporangeAI}{RGB}{255, 158, 128}
\definecolor{deeporangeAII}{RGB}{255, 110, 64}
\definecolor{deeporangeAIV}{RGB}{255, 61, 0}
\definecolor{deeporangeAVII}{RGB}{221, 44, 0}

\definecolor{brownOV}{RGB}{239, 235, 233}
\definecolor{brownI}{RGB}{215, 204, 200}
\definecolor{brownII}{RGB}{188, 170, 164}
\definecolor{brownIII}{RGB}{161, 136, 127}
\definecolor{brownIV}{RGB}{141, 110, 99}
\definecolor{brownV}{RGB}{121, 85, 72}
\definecolor{brownVI}{RGB}{109, 76, 65}
\definecolor{brownVII}{RGB}{93, 64, 55}
\definecolor{brownVIII}{RGB}{78, 52, 46}
\definecolor{brownIX}{RGB}{62, 39, 35}

\definecolor{grayOV}{RGB}{250, 250, 250}
\definecolor{grayI}{RGB}{245, 245, 245}
\definecolor{grayII}{RGB}{238, 238, 238}
\definecolor{grayIII}{RGB}{224, 224, 224}
\definecolor{grayIV}{RGB}{189, 189, 189}
\definecolor{grayV}{RGB}{158, 158, 158}
\definecolor{grayVI}{RGB}{117, 117, 117}
\definecolor{grayVII}{RGB}{97, 97, 97}
\definecolor{grayVIII}{RGB}{66, 66, 66}
\definecolor{grayIX}{RGB}{33, 33, 33}

\definecolor{bluegrayOV}{RGB}{236, 239, 241}
\definecolor{bluegrayI}{RGB}{207, 216, 220}
\definecolor{bluegrayII}{RGB}{176, 190, 197}
\definecolor{bluegrayIII}{RGB}{144, 164, 174}
\definecolor{bluegrayIV}{RGB}{120, 144, 156}
\definecolor{bluegrayV}{RGB}{96, 125, 139}
\definecolor{bluegrayVI}{RGB}{84, 110, 122}
\definecolor{bluegrayVII}{RGB}{69, 90, 100}
\definecolor{bluegrayVIII}{RGB}{55, 71, 79}
\definecolor{bluegrayIX}{RGB}{38, 50, 56}
\definecolor{bluegrayX}{RGB}{17, 23, 26}

\definecolor{myACMBlue}{cmyk}{1,0.1,0,0.1}
\definecolor{myACMYellow}{cmyk}{0,0.16,1,0}
\definecolor{myACMOrange}{cmyk}{0,0.42,1,0.01}
\definecolor{myACMRed}{cmyk}{0,0.90,0.86,0}
\definecolor{myACMLightBlue}{cmyk}{0.49,0.01,0,0}
\definecolor{myACMGreen}{cmyk}{0.20,0,1,0.19}
\definecolor{myACMPurple}{cmyk}{0.55,1,0,0.15}
\definecolor{myACMDarkBlue}{cmyk}{1,0.58,0,0.21}

\newcommand{\link}[1]{{\href{#1}{\color{blueVI}\textbf{\texttt{#1}}}}}
\newcommand{\linkhere}[2]{{\href{#1}{\color{blueVI}\textbf{#2}}}}

\newcommand{\mypar}[1]{\vspace{3pt}\noindent{}\textbf{{#1}}}

\newfloat{lstfloat}{htbp}{lop}
\floatname{lstfloat}{Code}

\lstdefinelanguage{TypeScript}{
  keywords={typeof, new, catch, function, return, null, catch, switch, var, if, in, while, do, else, case, break, let, const, await},
  keywordstyle=\color{lightblueVI}\bfseries,
  ndkeywords={class, export, boolean, throw, implements, import, this},
  ndkeywordstyle=\color{darkgray}\bfseries,
  identifierstyle=\color{black},
  sensitive=false,
  comment=[l]{//},
  morecomment=[s]{/*}{*/},
  commentstyle=\color{grayVI}\ttfamily,
  stringstyle=\color{pinkV}\ttfamily,
  morestring=[b]',
  morestring=[b]"
}

\newcommand{\figpart}[1]{\textcolor{myACMPurple}{#1}}

\newcommand{\headerspace}{\vspace{-1pt}}
\newcommand{\headerspacebottom}{\vspace{-1pt}}

\newcommand{\tool}{\textsc{MeMemo}}
\newcommand{\app}{\textsc{RAG Playground}}

\definecolor{soulorange}{RGB}{255, 212, 153}
\definecolor{soulgray}{RGB}{220, 220, 220}
\definecolor{soulgraylight}{RGB}{235, 235, 235}
\definecolor{soulred}{RGB}{252, 217, 218}
\definecolor{soulbluelight}{RGB}{208, 233, 253}
\definecolor{souldorangelight}{RGB}{254, 234, 212}
\colorlet{soulblue}{blueV!30}

\newcommand{\inlinefig}[2]{\raisebox{-1.5pt}{\includegraphics[height=#1pt]{figures/#2}}}

\definecolor{tagbordercolor}{rgb}{0.8, 0.8, 0.8}
\definecolor{tagbgcolor}{rgb}{0.9, 0.9, 0.9}
\definecolor{lightgray}{RGB}{247, 247, 247}
\definecolor{midgray}{RGB}{179, 179, 179}

\definecolor{tagbgcolor}{rgb}{1, 1, 1}

\definecolor{boxyellow}{RGB}{206, 171, 1}
\definecolor{boxgreen}{RGB}{14, 152, 136}
\definecolor{boxblue}{RGB}{77, 167, 223}

\setul{0.45ex}{0.2ex}

\begin{document}

\title{\tool{}: On-device Retrieval Augmentation for Private and Personalized Text Generation}

\author{Zijie J. Wang}
\orcid{0000-0003-4360-1423}
\email{jayw@gatech.edu}
\affiliation{%
  \institution{Georgia Institute of Technology}
  \city{Atlanta}
  \state{Georgia}
  \country{USA}
}

\author{Duen Horng Chau}
\orcid{0000-0001-9824-3323}
\email{polo@gatech.edu}
\affiliation{%
  \institution{Georgia Institute of Technology}
  \city{Atlanta}
  \state{Georgia}
  \country{USA}
}

\begin{abstract}
  Retrieval-augmented text generation (RAG) addresses the common limitations of large language models (LLMs), such as hallucination, by retrieving information from an updatable external knowledge base.
  However, existing approaches often require dedicated backend servers for data storage and retrieval, thereby limiting their applicability in use cases that require strict data privacy, such as personal finance, education, and medicine.
  To address the pressing need for client-side dense retrieval, we introduce \tool{}, the first open-source JavaScript toolkit that adapts the state-of-the-art approximate nearest neighbor search technique HNSW to browser environments.
  Developed with modern and native Web technologies, such as IndexedDB and Web Workers, our toolkit leverages client-side hardware capabilities to enable researchers and developers to efficiently search through millions of high-dimensional vectors in the browser.
  \tool{} enables exciting new design and research opportunities, such as private and personalized content creation and interactive prototyping, as demonstrated in our example application \app{}.
  Reflecting on our work, we discuss the opportunities and challenges for on-device dense retrieval.
  \tool{} is available at \link{https://github.com/poloclub/mememo}.
\end{abstract}



\begin{teaserfigure}
  \centering
  \includegraphics[width=470pt]{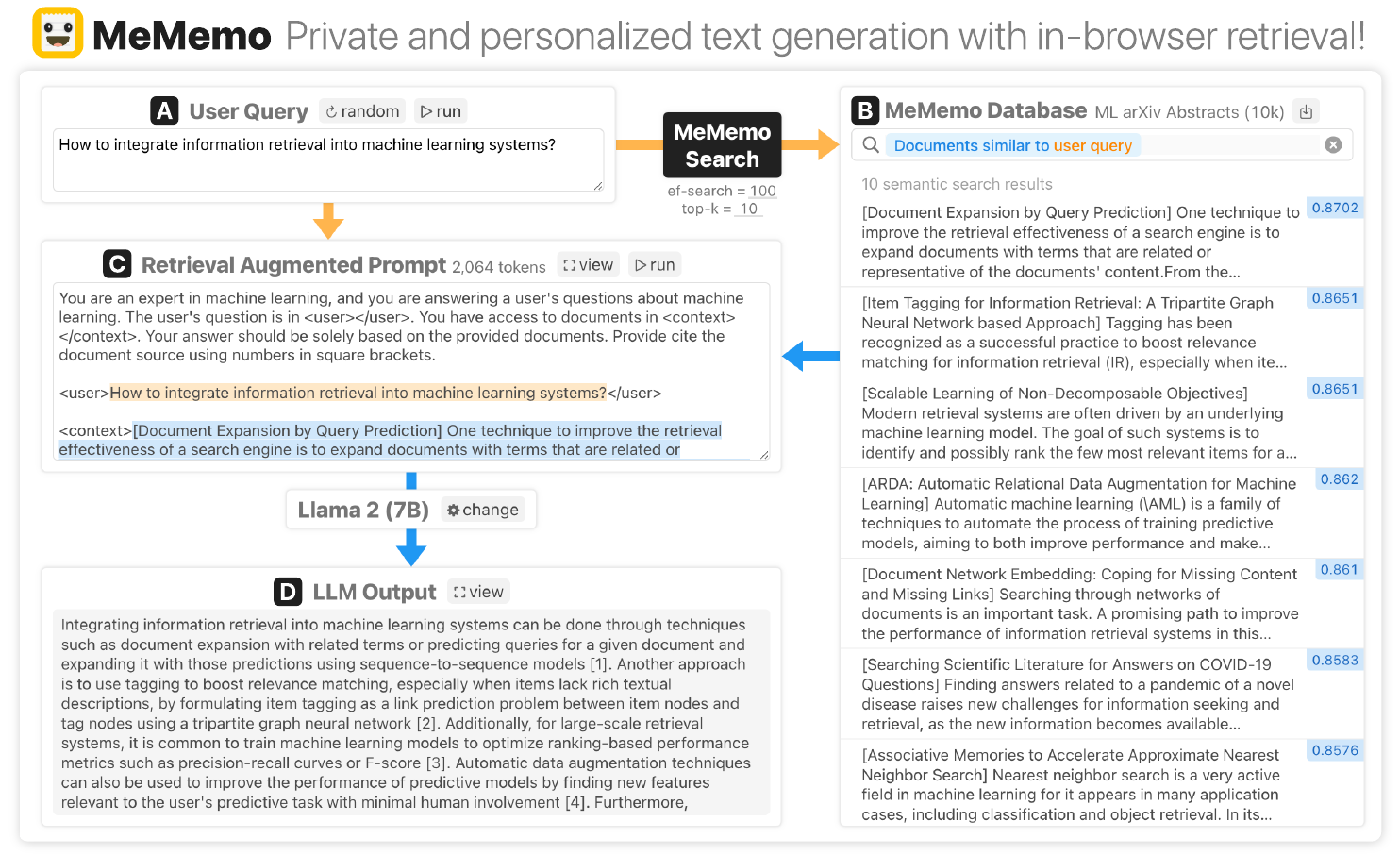}
  \vspace{-5pt}
  \caption{
    \tool{} is the first open-source JavaScript toolkit for in-browser dense neural retrieval.
    We demonstrate the capabilities of \tool{} by developing \app{} that enables AI developers to prototype retrieval-augmented text generation (RAG) apps locally in their browsers.
    With \app{}, developers can (A) enter various user queries, (B) search for semantically similar documents from an in-browser vector database, and (C) augment a text prompt with retrieved documents.
    (D) This allows developers to rapidly test if in-browser large language models generate more reliable responses to the query.
  }
  \Description{Teaser image for MeMemo.}
  \label{fig:teaser}
\end{teaserfigure}

\maketitle
\section{Introduction}

Retrieval augmented generation (RAG)~\cite{lewisRetrievalaugmentedGenerationKnowledgeintensive2020} with large language models (LLMs) has gained immense popularity from both practitioners and researchers, especially in applications such as domain-specific chatbots~\cite{semnaniWikiChatStoppingHallucination2023, princeOpportunitiesRetrievalTool2023}, code generation~\cite{soareDocuT5Seq2seqSQL2022, zhouDocPromptingGeneratingCode2023}, and interactive agents~\cite{hsiehToolDocumentationEnables2023, ruanTPTULargeLanguage2023}.
RAG can improve the accuracy and reliability of LLMs' generated text~\cite{shusterRetrievalAugmentationReduces2021}, by providing these models, such as GPT-4~\cite{openaiGPT4TechnicalReport2023} and Llama 2~\cite{touvronLlamaOpenFoundation2023}, with context information retrieved from an updatable and external knowledge base.
Compared to other techniques, such as fine-tuning~\cite{huLoRALowRankAdaptation2021} and prompt tuning~\cite{lesterPowerScaleParameterEfficient2021}, that aim to improve LLM's performance on new or specific domains, RAG is often favored by AI practitioners~\cite{martineauWhatRetrievalaugmentedGeneration2023} due to its ease of implementation, flexibility in maintenance, and superior performance~\cite{ovadiaFineTuningRetrievalComparing2024}.

However, current RAG systems rely on dedicated backend servers to store and retrieve external documents relevant to the user's query.
This is often achieved through nearest neighbor search using dense embedding vector representations of documents~\cite{balaguerRAGVsFinetuning2024, liLibVQToolkitOptimizing2023}.
The need for centralized backend servers limits the applicability of RAG in domains that prioritize data privacy, such as personal finance, education, and medicine~\cite[e.g.,][]{chungChallengesLargeLanguage2023, wutschitzRethinkingPrivacyMachine2023, fuchsbauerApproximateDistancecomparisonpreservingSymmetric2021, ghodratnamaAdaptingLLMsEfficient2023}.
Furthermore, implementing and hosting a vector storage and dense retriever pose additional challenges for AI novices and everyday LLM users~\cite{draxlerGenderAgeTechnology2023, zamfirescu-pereiraWhyJohnnyCan2023}, thereby increasing the barrier to entry for learning and applying RAG.

To address these pressing challenges, we present \tool{}, the first JavaScript toolkit that offloads vector storage and dense retrieval to the client---empowering a broader range of audiences to leverage cutting-edge retrieval techniques to enhance their LLM experiences.
Our work makes the following key \textbf{contributions:}

\begin{itemize}[topsep=5pt, itemsep=0mm, parsep=1mm, leftmargin=9pt]
      \item \textbf{\tool{}, the first scalable JavaScript library} that enables users to store and retrieve large vector databases directly in their browsers.
            Our toolkit adapts the state-of-the-art approximate nearest neighbor search Hierarchical Navigable Small World graphs~(HNSW)~\cite{malkovEfficientRobustApproximate2020} to the Web environment.
            By leveraging a novel prefetching strategy and modern Web technologies, such as IndexedDB and Web Workers, \tool{} empowers users to retrieve dense vectors with both privacy and efficiency~(\autoref{sec:design}).

      \item \textbf{\app{}, an example application of on-device dense retrieval.}
            We demonstrate the capabilities of \tool{} by developing \app{}~(\autoref{fig:teaser}), a novel client-side tool using on-device retrieval to enable interactive learning about RAG and rapid prototyping of RAG applications~(\autoref{sec:scenario}).
            We highlight the benefit of on-device retrieval regarding \textit{privacy}, \textit{ubiquity}, and \textit{interactivity}.
            Finally, we discuss the opportunities and challenges for future research on client-side retrieval augmentation and personalized text generation~(\autoref{sec:discussion}).
            \app{} is publicly accessible at \link{https://poloclub.github.io/mememo}.

      \item \textbf{An open-source\footnote{\tool{} code: \link{https://github.com/poloclub/mememo}} implementation} that lowers the barrier for researchers and developers to apply retrieval augmentation to improve text generation on the client side.
            We provide comprehensive documentation and an example application to help users use \tool{} to implement on-device retrieval augmentation across different Web environments.
            \tool{} is developed with minimal dependencies and TypeScript, a statically typed programming language, making it a maintainable and easy-to-use resource for the information retrieval community.
\end{itemize}

\noindent{}We hope our work will inspire the design, research, and development of on-device retrieval, enabling everyone to use text-generative models and other AI technologies more easily and privately.
\section{\tool{} in Action}
\label{sec:scenario}

We present two hypothetical usage scenarios, developing~(\autoref{sec:scenario:develop}) and using~(\autoref{sec:scenario:use}) \app{}, to demonstrate how researchers and practitioners can use \tool{} to easily develop client-side applications that take advantage of on-device RAGs.

\subsection{Developing In-browser RAG Tools}
\label{sec:scenario:develop}

\mypar{Motivations.}
Assume an example scenario where Mei, a machine learning (ML) consultant, is currently developing an LLM-based chatbot for a large design studio.
The chatbot's purpose is to assist new-hired designers in familiarizing themselves with the company's internal design systems and tools.
To ensure accurate and reliable responses, Mei integrates RAG into this onboarding chatbot.
This integration allows the responses to be grounded by relevant documentation, design documents, and code.
Initially, Mei uses Jupyter Notebooks~\cite{kluyverJupyterNotebooksaPublishing2016} to prototype the chatbot through prompt engineering in Python.
However, she realizes that this workflow is not ideal for collaborating with designers and introducing RAG to her clients.
This is because many of the collaborators and stakeholders are not experienced in programming and setting up notebook environments.
Therefore, Mei decides to develop \app{}~(\autoref{fig:teaser}), a web-based no-code RAG prototyping tool.
This tool will enable her collaborators, who come from diverse backgrounds, to easily access and prototype RAG features for their chatbot through their web browsers.

\begin{lstfloat}[tb]
  \renewcommand{\lstlistingname}{Code}
  \begin{lstlisting}[belowskip=-10pt, language=TypeScript, caption=Example TypeScript code that uses \tool{} to create an HNSW index and search for k-nearest neighbors.,label=lst:creation]
import { HNSW } from 'mememo';

// Creating a new index
const index = new HNSW({ distanceFunction: 'cosine' });

// Inserting elements, keys: string[], values: number[][]
await index.bulkInsert(keys, values);

// Find k-nearest neighbors, query: number[], k: number
// keys: string[], distances: number[]
const { keys, distances } = await index.query(query, k);
\end{lstlisting}
\end{lstfloat}

\mypar{Vector storage and retrieval with \tool{}.}
Mei uses \tool{}, a JavaScript library, to enable dense vector storage and retrieval directly in the browser.
By installing the library with a single command \texttt{npm install mememo}, Mei can easily import it into her web app, regardless of her web development stack (e.g., JavaScript, TypeScript, React~\cite{facebookReactLibraryWeb2013}, Svelte~\cite{harrisSvelteCyberneticallyEnhanced2016}, or Lit~\cite{googleLitSimpleFast2015}).
With just a few lines of code~(\autoref{lst:creation}), Mei can create an HNSW vector index~\cite{malkovEfficientRobustApproximate2020} and efficiently search through millions of embedding vectors entirely within her browser.
Mei also uses \tool{}'s \texttt{exportIndex()} and \texttt{loadIndex()} functions to export an index she has created into persistent local storage or as a JSON file.
This allows her collaborators to quickly load the HNSW index without the need to recreate it every time they use \app{}.

\mypar{Smooth integration with existing Web ML technologies.}
\label{sec:scenario:develop:integrate}
Mei seamlessly integrates \tool{} with other Web ML technologies.
For example, she uses IndexedDB, a client-side key-value browser storage, to store the raw documents.
Using the same keys, Mei creates the HNSW index with \tool{}.
Then, Mei uses FlexSearch~\cite{wilkerlingFlexSearchNextGenerationFull2019} to implement fast full-text lexical search in the browser.
To enable semantic search, Mei first uses GTE-Small~\cite{liGeneralTextEmbeddings2023} to encode all documents into dense vectors with 384 dimensions in Python with SentenceTransformers~\cite{reimersSentenceBERTSentenceEmbeddings2019}.
For encoding the user's query~(\autoref{fig:teaser}\figpart{A}), Mei uses ONNX~\cite{baiONNXOpenNeural2019} and Transformer.js~\cite{lochnerTransformersJsStateoftheart2023} to run the same GTE-Small model in the browser.
After augmenting a text prompt with retrieved documents~(\autoref{fig:teaser}\figpart{C}), Mei runs the prompt with open-source LLMs, such as LLama 2~\cite{touvronLlamaOpenFoundation2023} and Phi 2~\cite{abdinPhi2SurprisingPower2023}, in the browser through Web LLM~\cite{teamMLCLLM2023}.
By combining \tool{} with existing Web ML technologies, Mei quickly develops \app{} and shares it with her collaborators.
With this tool, Mei's team has made great progress as all stakeholders with diverse backgrounds can easily experiment with different user queries and prompts to improve their onboarding chatbot.

\subsection{Prototyping with \app{}}
\label{sec:scenario:use}

\mypar{Motivations.}
Robaire, a graduate student studying human-computer interaction, is designing an interactive visualization tool to assist researchers in brainstorming and literature review.
After discovering RAG online, Robaire becomes interested in integrating it into his prototype.
The objective is to allow users to input a large corpus of academic papers and use natural language queries to discover related papers and visualize the connections between them.
Since Robaire has never implemented RAG before, he turns to \app{} to learn about the concept and prototype for his tool.

\setlength{\columnsep}{6pt}%
\setlength{\intextsep}{0pt}%
\begin{wrapfigure}{R}{0.17\textwidth}
  \vspace{-1pt}
  \centering
  \includegraphics[width=0.17\textwidth]{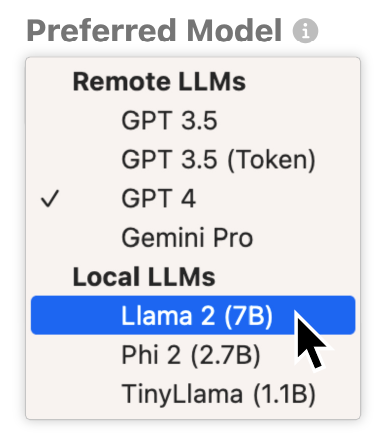}
\end{wrapfigure}
\mypar{Learning and experimenting with RAG.}
After opening \app{} in the browser, Robaire creates a \tool{} database~(\autoref{fig:teaser}\figpart{B}) by uploading a JSON file containing the abstracts of 120k arXiv ML papers and 384-dimensional embeddings of the abstracts.
Robaire then pretends to be his end-users and types in a natural language query in the \textit{User Query View}, such as ``how to integrate information retrieval into ML?''~(\autoref{fig:teaser}\figpart{A}).
In addition, he writes a simple system prompt template in the \textit{Prompt View}~(\autoref{fig:teaser}\figpart{C}) with placeholders \texttt{\{\{user\}\}} and \texttt{\{\{context\}\}}.
After clicking the~\inlinefig{9}{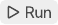} button, Robaire sees 10 relevant paper abstracts with their Cosine distances highlighted in the \textit{Database View}~(\autoref{fig:teaser}\figpart{B}).
He also finds that the two placeholders in the \textit{Prompt View} are replaced with the user query and relevant documents.
Robaire then sees the LLM's output in the \textit{Output View}~(\autoref{fig:teaser}\figpart{D}).
Finding the output helpful and grounded by the documents retrieved by \tool{}, Robaire experiments with more prompts and both remote and local LLMs~(e.g., GPT 4 and Llama 2 shown in the figure above) in \app{} and gains a better understanding of RAG.
This increased understanding gives him more confidence to implement RAG in his tool.\looseness=-1 %
\section{\tool{} Design and Implementation}
\label{sec:design}

\tool{} is the first JavaScript toolkit that enables dense retrieval in the browser.
To enable fast and reliable retrieval for RAG, our tool adapts the state-of-the-art approximate nearest neighbor search technique HNSW~(\autoref{sec:design:hnsw}).
\tool{} leveraging modern and native Web technologies, such as IndexedDB and Web Workers~(\autoref{sec:design:optimize}), to optimize for browser environments.
To help researchers and developers adopt \tool{}, we have open-sourced it and provided detailed documentation, tutorial, and an example application~(\autoref{sec:design:open}).

\subsection{Adapting HNSW}
\label{sec:design:hnsw}

HNSW is a state-of-the-art approximate k-nearest neighbor search technique introduced by \citeauthor{malkovEfficientRobustApproximate2020}.
It is inspired by the greedy graph routing used in navigable small world networks~\cite{kleinbergNavigationSmallWorld2000,bogunaNavigabilityComplexNetworks2009} and the stochastic hierarchical structure in 1D probabilistic skip list~\cite{pughSkipListsProbabilistic1990}.
HNSW uses a multilayered graph structure to connect high-dimensional dense vectors.
During the insertion process, each new element is assigned a layer level at random, determining its position within the graph's multi-layered hierarchy.
The insertion process involves finding the element's closest neighbors, starting from the top layer and working downwards using a greedy search approach.
When searching for the nearest neighbors of a query element, the algorithm follows a similar procedure.
It starts from the top layer and uses the connections established during the insertion phase to guide its search downwards.

We use HNSW as the approximate nearest neighbor search technique in \tool{} because it is the state-of-the-art regarding construction and query efficiency~\cite{malkovEfficientRobustApproximate2020}.
Additionally, HNSW has gained immense popularity among retrieval and AI practitioners and has been integrated into popular retrieval and RAG Python toolkits such as FAISS~\cite{douzeFaissLibrary2024}, Pyserini~\cite{linPyseriniPythonToolkit2021}, PGVector~\cite{kanePgvectorOpensourceVector2021}, and LangChain~\cite{chaseLangChainBuildingApplications2022}.
Our goal with \tool{} is to seamlessly integrate into users' existing workflows and preferences, providing a smooth and familiar experience when developing in-browser retrieval applications.

\subsection{Optimizing for the Browsers}
\label{sec:design:optimize}

\mypar{Memory management.}
Memory management is one of the main challenges for developing in-browser toolkits.
Depending on the device and browser, a webpage tab might have a RAM limit as low as 256MB~\cite{maitreTotalCanvasMemory2018}.
This means that without considering any other memory usage on a webpage, it can store at most 83k 384-dimensional vectors in RAM.
Additionally, for security reasons, browsers do not allow access to the operating system's file systems, so \tool{} cannot directly store data in the user's disk.
To overcome these challenges, \tool{} leverages IndexedDB~\cite{mdnIndexedDBAPIWeb2021}, a cross-browser key-value storage that can use up to 80\% of the client's disk size~\cite{mdnStorageQuotasEviction2023}.
In IndexedDB, \tool{} stores all vector values, while only keeping the keys and HNSW graphs in the RAM.

\mypar{Prefetching for efficient data access.}
While IndexedDB addresses the memory constraints in the browser, reading or writing a large amount of data to IndexedDB with consecutive transactions is extremely slow~\cite{rxdbWhyIndexedDBSlow2021}.
Dexie.js~\cite{fahlanderDexieJsMinimalistic2021} introduces techniques for fast batched read and write to IndexedDB.
However, the HNSW construction process requires consecutive reads and writes of vector values, as the algorithm relies on the previously constructed index for finding good neighbors~\cite{mendel-gleasonParallelisingHNSW2024,malkovEfficientRobustApproximate2020}.
To address this challenge, \tool{} introduces a prefetching mechanism.
When inserting multiple elements, \tool{} first uses a batched write to store all vectors in IndexedDB.
During construction and search, \tool{} maintains a cache of $p$ vector values in RAM.
If it needs to read a vector value that is not in the cache, \tool{} prefetches $p$ neighbors of that element on the current graph layer from IndexedDB to RAM.
This mechanism reduces the number of IndexedDB transactions.
The parameter $p$ is automatically determined by the vector dimension and can be configured by users.

\subsection{Open-source and Easy to Use}
\label{sec:design:open}

To help researchers and developers easily adopt \tool{}, we open source our implementation and design APIs similar to popular HNSW Python libraries~\cite[e.g.,][]{malkovEfficientRobustApproximate2020,douzeFaissLibrary2024,zhuEkzhuDatasketchMinHash2016}.
Users can easily configure all HNSW parameters, such as $M$ (the number of neighbors a graph node can have) and $\mathit{efConstruction}$ (the number of nodes to search during construction).
With just a few lines of code~(\autoref{lst:creation}), users can quickly implement dense retrieval in web browsers using \tool{}.
We provide detailed documentation and tutorials.
Additionally, we offer an open-source example application \app{} that demonstrates the integration of \tool{} with existing Web ML technologies~(\autoref{sec:scenario:develop}).
\app{} also shows how to use \tool{} with modern Web APIs, including Web Workers~\cite{mdnWebWorkersAPI2023} to prevent blocking the main thread and Streams API~\cite{mdnStreamsAPIWeb2023} for creating an HNSW index incrementally with small network-received chunk.
\tool{} is published in the popular  \linkhere{https://www.npmjs.com/package/mememo}{Web package repository \texttt{npm} Registry}, and can be easily installed and used in both browser and Node.js~\cite{dahlNodeJsOpensource2009} environments.
\headerspace{}
\section{Related Work}
\label{sec:related}
\headerspacebottom{}

\mypar{Retrieval-augmented text generation.}
There has been a long history of using information retrieval to enhance text generation, such as developing language models through retrieval~\cite{lavrenkoRelevanceBasedLanguage2001}, using a retrieve-and-edit framework to improve code generation~\cite{hashimotoRetrieveandeditFrameworkPredicting2018}, and incorporating knowledge graphs to enhance language representation in language models~\cite{zhangERNIEEnhancedLanguage2019}.
The concept of RAG was popularized by \citeauthor{lewisRetrievalaugmentedGenerationKnowledgeintensive2020}, who introduced a model that combines a dense passage reliever and sequence-to-sequence models.
More recent approaches~\cite[e.g.,][]{neelakantanTextCodeEmbeddings2022, quPassageRetrievalOutsideKnowledge2021, izacardLeveragingPassageRetrieval2021, cuconasuPowerNoiseRedefining2024} use pre-trained embedding models to encode external documents as dense vectors and retrieve relevant documents using dense retrievers such as HNSW~\cite{malkovEfficientRobustApproximate2020}, PQ~\cite{jegouProductQuantizationNearest2011}, and FAISS~\cite{douzeFaissLibrary2024}.
\tool{} builds upon these works and extends RAG to the client side for more private and personalized text generation.

\mypar{On-device retrieval and machine learning.}
Traditional retrieval and machine learning (ML) systems are typically deployed on remote servers, and their outputs are sent to client devices.
However, there has been a recent surge of interest in deploying ML models directly on edge devices in the pursuit of private, ubiquitous, and interactive ML experiences.
Tools such as TensorFlow.js~\cite{smilkovTensorFlowJsMachine2019}, ONNX~\cite{baiONNXOpenNeural2019}, MLC~\cite{teammlcMLCLLM2023, chenTVMAutomatedEndtoEnd2018}, and Core ML~\cite{appleCoreMLIntegrate2017} have significantly reduced the barriers to running complex ML models in browsers and mobile devices.
Researchers have proposed various on-device systems, including information retrieval~\cite{kamvarComputersIphonesMobile2009, lamGPUbasedPrivateInformation2023}, recommender systems~\cite{gongEdgeRecRecommenderSystem2020, xiaEfficientOnDeviceSessionBased2023}, prediction explanation~\cite{wangInterpretabilityThenWhat2022,wangGAMCoachInteractive2023,wangWebSHAPExplainingAny2023}, speech recognition~\cite{macoskeyAmortizedNeuralNetworks2021,macoskeyLearningNeuralDiff2021}, translation~\cite{tanDynamicMultiBranchLayers2022}, and writing assistants~\cite{wangWordflowSocialPrompt2024}.
Our tool contributes to the growing body of on-device ML research by introducing the first adaptation of dense retrieval to browsers.

\section{Discussion and Future Work}
\label{sec:discussion}

Reflecting on our development of \tool{}, we highlight the opportunities and challenges for in-browser dense retrieval.

\mypar{Opportunities.}
Enabling dense retrieval and RAG in browsers offers significant advantages regarding \textit{privacy}, \textit{ubiquity}, and \textit{interactivity}.
With the browser's ubiquity, \tool{} is accessible on various devices, including laptops, mobile phones, and IoT appliances like smart refrigerators.
Future research directions include:

\begin{itemize}[topsep=3pt, itemsep=2pt, parsep=2pt, leftmargin=9pt]
      \item \textbf{Intelligent personal information management.}
            There is a large body of research on collecting all of one's personal information into a searchable database~\cite[e.g.,][]{freemanLifestreamsStorageModel1996, caiPersonalInformationManagement2005, bellPersonalDigitalStore2001, chauWhatWhenSearch2008, kieselWASPWebArchiving2018}.
            Researchers can leverage on-device dense storage and retrieval to design browser extensions that automatically and privately encode and store a user's visited web pages, photos, and academic papers.
            These extensions can serve as an intelligent ``second brain''~\cite{forteBuildingSecondBrain2022} to help users capture and review knowledge.

      \item \textbf{Private and personalized content creation.}
            If users maintain a personal vector database in browsers, content creators, such as book writers, can use on-device RAG to tailor their content privately based on readers' preferences and reading history.

      \item \textbf{Interactive RAG prototyping.}
            Future researchers can enhance the design of \app{} to improve interactive RAG prototyping experience, such as supporting collaborative prompt editing~\cite{fengCoPromptSupportingPrompt2023} and interactive embedding visualizations~\cite{wangWizMapScalableInteractive2023}.
\end{itemize}

\mypar{Challenges.}
Due to limited computation resources in browsers, \tool{} is slower than heavily optimized libraries like \texttt{HNSWLIB}~\cite{malkovEfficientRobustApproximate2020} in terms of index creation and search.
In Chrome on a 64GB RAM MacBook, it took about 94 minutes to insert 1 million 384-dimensional vectors ($M$=5, $\mathit{efConstruction}$=20).
However, querying this index with 1M items is still performed in real time.
Future researchers can optimize in-browser dense retrieval further by implementing parallelization and smarter prefetching techniques.

\mypar{Conclusions.}
We present \tool{}, an open-source library that enables in-browser dense retrieval using HNSW and modern Web technologies.
We introduce \app{}, a novel client-side RAG prototyping tool to demonstrate the capabilities of \tool{}.
We hope \tool{} to be an easy-to-use resource for the information retrieval and ML community, inspiring future research and development of on-device retrieval and RAG applications.

\clearpage{}
\bibliographystyle{ACM-Reference-Format}
\balance
\bibliography{24-mememo}

\end{document}